\documentclass{sigchi-ext}
\usepackage[T1]{fontenc}
\usepackage{textcomp}
\usepackage[scaled=.92]{helvet} 
\usepackage{graphicx} 
\usepackage{balance}  
\usepackage{booktabs} 
\usepackage{ccicons}  
\usepackage{ragged2e} 

\usepackage[all]{nowidow}

\usepackage[utf8]{inputenc} 

\def\plaintitle{Considering Gut Biofeedback for Emotion Regulation}
\def\plainauthor{Jelena Mladenović}
\def\plainkeywords{Electrogastrography; Physiological sensors; Gut Brain axis; Biofeedback}

\title{\plaintitle}

\numberofauthors{1}
\author{%
  \alignauthor{%
    \textbf{Jelena Mladenović}\\
    \affaddr{Inria, France}\\
    \email{jelena.mladenovic@inria.fr} } 
 }

\definecolor{linkColor}{RGB}{6,125,233}

\hypersetup{%
  pdftitle={\plaintitle},
  pdfauthor={\plainauthor},
  pdfkeywords={\plainkeywords},
  bookmarksnumbered,
  pdfstartview={FitH},
  colorlinks,
  citecolor=black,
  filecolor=black,
  linkcolor=black,
  urlcolor=linkColor,
  breaklinks=true,
}

\copyrightinfo{Permission to make digital or hard copies of part or all of this work for personal or classroom use is granted without fee provided that copies are not made or distributed for profit or commercial advantage and that copies bear this notice and the full citation on the first page. Copyrights for third-party components of this work must be honored. For all other uses, contact the Owner/Author.\\
Copyright is held by the owner/author(s).\\
{\emph{UbiComp/ISWC'18 Adjunct}}, October 8-12, 2018, Singapore, Singapore.\\
ACM 978-1-4503-5966-5/18/10. \\
https://doi.org/10.1145/3267305.3267686}

\begin{document}

\teaser{
 \centering
 \includegraphics[width=1\textwidth]{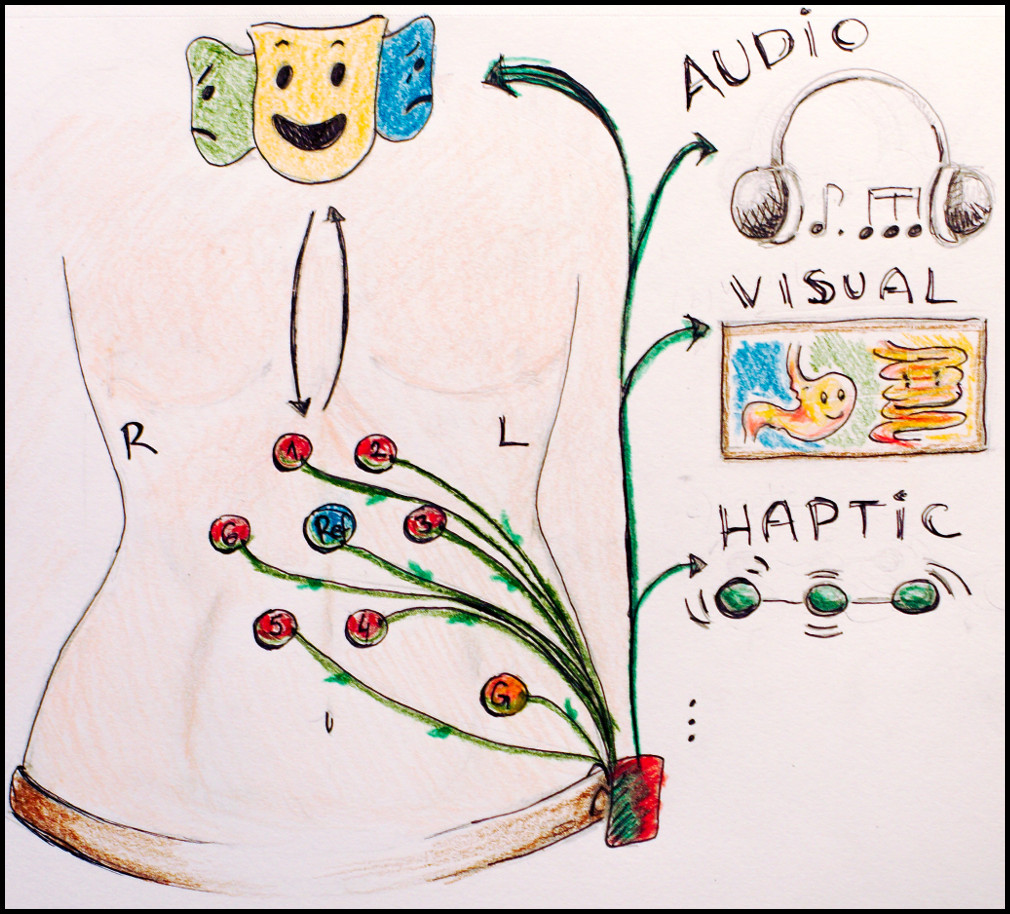}
 \caption{Depiction of a potential gut biofeedback for regulating emotion through various modalities out of which we expose audio, visual and haptic. The electrode positioning is from \protect\cite{gharibans2018}. \label{fig:teaser}}}

\maketitle

\RaggedRight{} 


\begin{abstract}
Recent research in the enteric nervous system, sometimes called the second brain, has revealed potential of the digestive system in predicting emotion. Even though people regularly experience changes in their gastrointestinal (GI) tract which influence their mood and behavior multiple times per day, robust measurements and wearable devices are not quite developed for such phenomena. However, other manifestations of the autonomic nervous system such as electrodermal activity, heart rate, and facial muscle movement have been extensively used as measures of emotions or in biofeedback applications, while neglecting the gut.  We expose electrogastrography (EGG), i.e., recordings of the myoelectric activity of the GI tract, as a possible measure for inferring human emotions. In this paper, we also wish to bring into light some fundamental questions about emotions, which are often taken for granted in the field of Human Computer Interaction, but are still a great debate in the fields of cognitive neuroscience and psychology.

\end{abstract}



\section{Introduction}

Recent developments in Human Computer Interaction (HCI), and physiological and affective computing  brought to light the necessity for wearable and robust physiological sensors.
So far, using physiological sensors a person can: (1) consciously monitor/regulate  their  bodily  functions through biofeedback for well-being  \cite{McKee2008},  (2) (un)consciously  adapt  an  environment or task, which can for instance increase immersion in gaming \cite{van2013},  or  (3) consciously  manipulate  an  external device  with only  physiological (neural)  activity, as in active Brain-Computer Interfaces, to control wheelchairs or for communication for example\cite{wolpaw2002brain}.
Measures of  electrodermal activity (EDA), cardiac function, facial muscles activity, and respiration have been used frequently to assess emotional states \cite{mayer2000}. Nowadays there are wearable devices developed for measuring EDA and  heart rate, such as the Empatica E4 smartwatch. Remarkably however, the gastrointestinal system has often been neglected by affective research. Even though humans regularly experience having a "gut feeling" or "butterflies in the stomach", they often overlook the importance of such phenomenon as an actual physiological process. However, studies have shown that indeed the gut could have an important role in affective disorders \cite{bennett1998}. Still, non-invasive, robust physiological measurements or wearable devices for such phenomena are not yet developed. The possibility of assisting users in regulating the internal processes of the gut, and thus regulating the emotions that arise with such physiological processes are not yet taken seriously into consideration. 

In this paper we briefly explain what the gut signal is, and the usefulness of such modality for inferring and regulating emotions, using a biofeedback. We also tackle some fundamental questions about emotions which are often taken lightly in the HCI community.

\section{Gastro-Intestinal tract}
The gastro-intestinal (GI) tract comprises of the mouth, esophagus, stomach and intestines. The GI tract has a bidirectional communication with the Central Nervous System (CNS) through the sympathetic and parasympathetic systems \cite{sudo2004}, thus researchers often refer to the gut-brain axis. The GI tract is governed by the enteric nervous system which can act independently from the CNS and contains over 500 million nerves, which is why it is also called the "second brain". Moreover, today there has been many interest in the gut microbiota or microorganisms that inhabit the gut and have shown to have a role in the stress regulation in mice \cite{sudo2004}.

The electrogastrogram (EGG) is a reliable and noninvasive method of recording gastric myoelectrical activity \cite{nelsen1968}. The gastric myoelectrical activity paces the contraction of the stomach. The normal frequency of the electrogastric wave is 3 cycles per minute (cpm), and is termed normogastria \cite{koch2004}. It is worth nothing that amplifiers typically used for electroencephalography (assessing brain activity) have shown to be equally useful for EGG, for example in \cite{gharibans2018} using an affordable and open-source device, OpenBCI. 
Recent studies showed that EGG could be a valuable measure of emotion \cite{vianna2006}. 
Individuals often report a "nervous stomach" for too frequent contractions (tachygastria, 4–9 cpm) during stressful experiences \cite{vujic2018}. Participants reacted with tachygastria during horror movies, but a reduced frequency of gastric waves during a relaxation session \cite{yin2004}. It is also shown that gastric slow waves can be useful for predicting the experience of disgust \cite{harrison2010embodiment}.

Individuals clearly react emotionally with their gut, as well as the gut influences their emotions. As such, we advocate that it could be interesting to propose biofeedback specifically aimed at regulating a "nervous stomach".

\section{Biofeedback for gut awareness}

Biofeedback is a system that externalizes one's internal bodily activity, for example in visual, audio or haptic modalities. It assists people to be aware of their internal processes or physiological activity, as a technique of interoception, known to be beneficial for well-being \cite{Farb2015}.
Notice that biofeedback is built under the assumption that being aware of one's physiological processes creates or modulates an emotion. In other words, the perception of physiological changes contributes to the content of conscious experiences of emotion \cite{tsuchiya2007emotion}. Biofeedback thus externalizes such phenomena and enables people to consciously examine and regulate their internal states and their experience of emotions. 
As the gut clearly has an important role in human emotion, we believe it could be beneficial to build an EGG wearable device which could record and process feedback to one's gut contractions, as depicted in Figure \ref{fig:teaser}. Interestingly, the use of biofeedback could also expose the relationship between experiencing bodily activity and experiencing an emotion. In experiments where people are given a fake biofeedback to manipulate their emotions toward images of individuals, the perception of external audio stimuli dominated over their autonomic perception \cite{woll1979effects}. This leads us to ask whether the perceived physiological process is more important than the actual one.

\section{Relation between physiology and emotion}

Sympathetic nervous system, governing the fight or flight mechanisms, influences sweat secretion, increases heart rate, constricts blood vessels in gastrointestinal organs or inhibits contractions in the digestive tract, and much more. These physiological changes are recognized as measures of emotion and expressed as stress, anxiety, fear etc. This assumption follows the James' theory \cite{james1884emotion} in which feeling (emotion experience) exists due to physiological changes in one's own body. James argued that seeing a fearful stimulus would first trigger emotional responses (increases in sympathetic activity), and that the perception of these physiological changes would form the basis for our conscious experience of emotion.
Today, in affective neuroscience, the James theory is revised and updated, e.g. acknowledging the role of emotions in decision-making \cite{bechara2000emotion}; or distinguishing "the conscious experience of an emotion (feeling), its expression (physiological response), and semantic knowledge about it (recognition)" \cite{tsuchiya2007emotion}.
Taking more often into consideration the role of the GI tract might help to reconcile antagonist views of emotion. For example, in \cite{johnsen2009} authors described the dissociation between the autonomic response and affect through the study of patients with brain lesions. In this experiment,  patients without automonic responses would not sweat but would still be able to experience emotions related to music excerpts, while patients with different lesions, incapable of judging music, displayed EDA responses. As such, without a link between physiology and emotions, authors "opposed" James' theory. Nevertheless, we believe, as the enteric nervous system can function independently from the autonomic system, it could be that the physiology still contributed to the emotional perception of music.

\section{Conclusion}

With this paper we hope to foster discussions among HCI practitioners about the study of gut signals. To discover further how the body contributes to the experience of emotion and \textit{vice versa}, it can be useful to include EGG as an additional tool for emotion recognition. Also, affordable and mobile biosignal amplifiers could enable the creation of a new biofeedback mechanism, in which individuals could learn how to regulate their emotion related to the gut.

\section{Acknowledgment}
I wish to thank Jérémy Frey and Angela Vujić for insightful discussions and for proofreading this paper. 

\balance{}
\bibliographystyle{SIGCHI-Reference-Format}
\bibliography{biblio}

\end{document}